\documentclass[a4paper,fleqn]{cas-dc}

\usepackage[numbers]{natbib}

\usepackage{siunitx}
\usepackage{amsmath}
\usepackage{booktabs}
\usepackage{multirow}
\usepackage{graphicx}
\usepackage{subcaption} 

\begin{document}
\let\WriteBookmarks\relax
\def\floatpagepagefraction{1}
\def\textpagefraction{.001}

\shorttitle{Fused Silica Activation Cherenkov Detector}
\shortauthors{N. Kaneshige et~al.}

\title[mode=title]{Fused-Silica Activation Cherenkov Detector for Pulsed D--T Fusion Yields}

\author[1]{N. Kaneshige}\cormark[1]
\ead{na167k@helionenergy.com}
\author[1]{S. Alawabdeh}
\author[1]{W. Hennig}
\author[1]{D. Cech}
\author[1]{M. Hua}
\author[1]{R. Grazioso}

\affiliation[1]{organization={Helion Energy Inc.},
  city={Everett, WA},country={USA}}

\cortext[cor1]{Corresponding author.}

\begin{abstract}
We demonstrate a compact, non-toxic, low-cost neutron-yield diagnostic for pulsed D--T fusion machines that uses an undoped fused-silica (SiO\textsubscript{2}) rod as both activation target and Cherenkov radiator. D--T neutrons (14.1~MeV) activate \textsuperscript{28}Si and \textsuperscript{16}O to produce short-lived \textsuperscript{28}Al (\(T_{1/2}=134~\text{s}\)) and \textsuperscript{16}N (\(T_{1/2}=7.13~\text{s}\)); ensuing \(\beta^-\) particles exceed the Cherenkov threshold and generate UV--visible light detected by a fast photomultiplier tube. A \SI{6}{in\ length}\,\(\times\)\,\SI{1}{in\ diameter} SiO\textsubscript{2} rod is optically coupled and read out with a CAEN DT5730 digitizer with digital pulse processing firmware operating in list mode; the post-pulse count rate is fit with fixed \(^{16}\)N, \(^{28}\)Al half-lives plus background terms to infer fluence. Testing at the ZEUS D–T Dense Plasma Focus (DPF) established a reference calibration and close agreement with a praseodymium-calibrated silver activation detector. Testing near a D--D DPF shows no activation signal, confirming D--T selectivity. The approach enables pulse-to-pulse yield measurements in minutes following a pulse and is being used on Helion Energy's 7th fusion prototype: Polaris.
\end{abstract}

\begin{highlights}
\item D--T-selective activation--Cherenkov detector using undoped fused silica
\item Measures single-shot pulsed D--T fluences
\item Shot-to-shot yield in minutes via multi-exponential decay fits
\item Insensitive to D--D neutrons (\SI{2.45}{MeV}) and most gamma backgrounds
\end{highlights}


\maketitle

\section{Introduction}
Resolving the neutron yield for each individual pulse is central to operating and optimizing pulsed fusion machines. Traditional activation foils provide robust absolute yield but require handling and delayed readout, extending turnaround times to hours. Prompt scintillation detectors can be fast but are susceptible to saturation, electromagnetic interference, and gamma/neutron discrimination challenges in pulsed environments. We report a fused-silica activation Cherenkov detector that integrates activation material and radiator into a single, inert medium, producing \(\beta\)-induced Cherenkov signals that are easy to digitize and fit for D--T neutron fluences. This detector is fielded at Helion Energy on the Polaris fusion energy prototype and regularly calibrated against zirconium neutron-activation foils counted on a high-purity germanium (HPGe) detector with NIST-traceable efficiency calibration~\cite{Helion}.

\begin{figure}[b]
\centering
\includegraphics[width=\columnwidth,clip,trim=3.3cm 0 3cm 0]{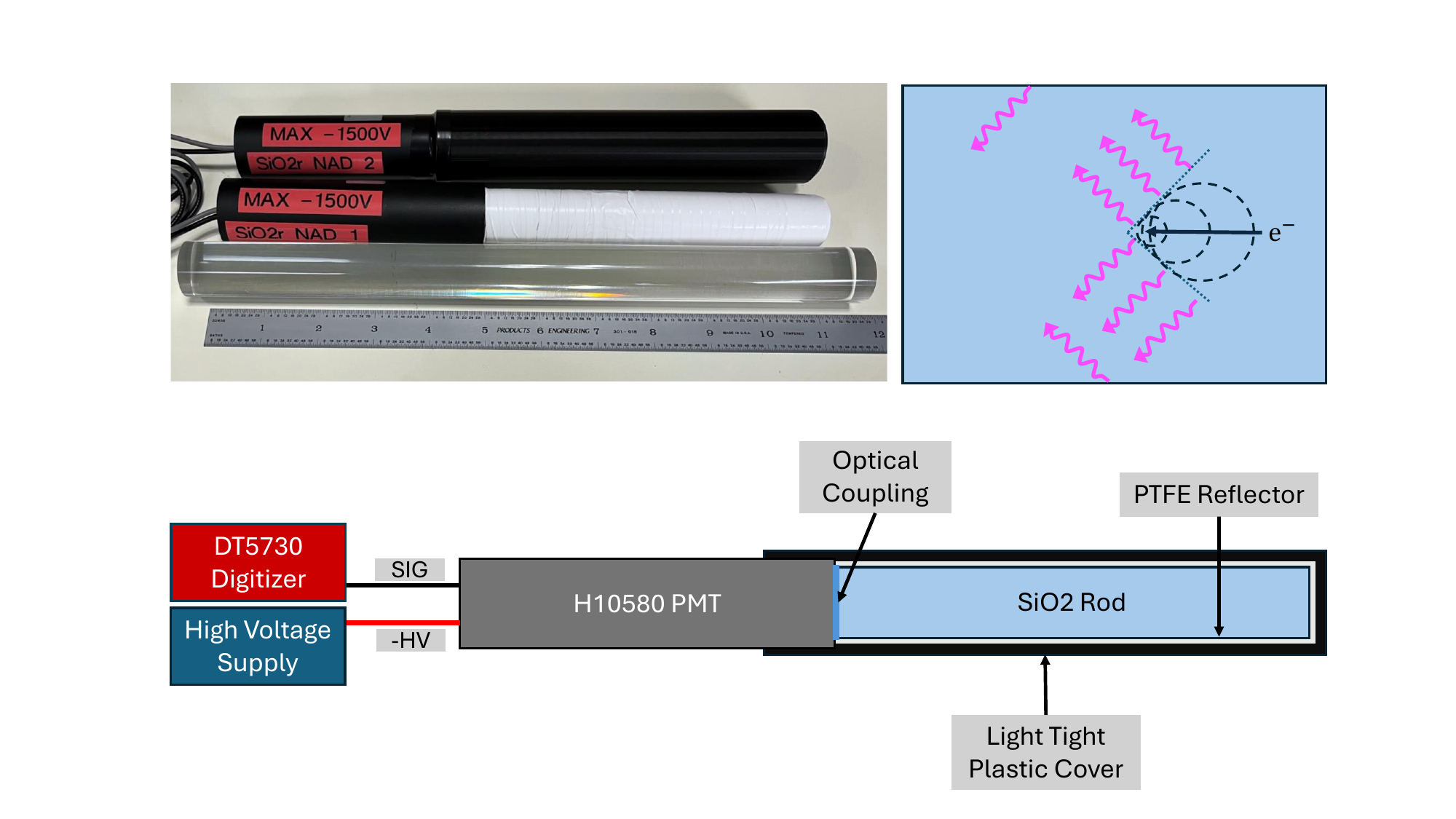}
\caption{Photo and diagram of the fused-silica activation Cherenkov detector. A SiO\textsubscript{2} rod is optically coupled to a Hamamatsu H10580 PMT. D--T neutrons induce \(^{28}\mathrm{Si(n,p)}^{28}\mathrm{Al}\) and \(^{16}\mathrm{O(n,p)}^{16}\mathrm{N}\); ensuing \(\beta^-\) particles generate Cherenkov photons detected by the PMT and are digitized by a CAEN DT5730 digitizer. }
\label{fig:detector}
\end{figure}

\section{Background and Related Work}
Activation diagnostics for pulsed fusion machines exploit threshold reactions and short-lived products to convert an instantaneous neutron fluence into decay rates with known activities and lifetimes. Materials such as Ag, In, Cu, Zr, and Pr provide robust absolute yield and energy selectivity, but they require removal and counting on separate HPGe/NaI systems, extending turnaround to hours ~\cite{Cooper2001,Meehan2010,Hahn2014}.

To increase immediacy of fusion yield results, hybrid activation–scintillation detectors have been developed. The silver-activation detector of Slaughter and Pickles coupled layers of silver foils to plastic scintillators for sensitivity on pulsed sources through silver's thermal neutron activation and decay~\cite{Slaughter1979}. Other materials target different thresholds but add practical burdens: beryllium–plastic scintillator stacks rely on \(^{9}\mathrm{Be(n,\alpha)}{}^{6}\mathrm{He}\) (\(T_{1/2}\!\approx\!0.8\) s) decays detected in adjacent plastic scintillators, and arsenic devices impregnate As into epoxies to utilize \(^{75}\)As(n,n\('\))\(^{75\mathrm{m}}\)As; both introduce toxicity and construction complexity (multilayer stacks or hazardous encapsulation) that raise cost and limit deployability~\cite{Ruiz2019,Jacobs1983}. Non-toxic systems have been explored such as the “O-probes” for the Z-facility leveraging \(^{16}\mathrm{O(n,p)}{}^{16}\mathrm{N}\) in inert layers of glass stacked with plastic scintillators~\cite{Mangan2024}.

At the National Ignition Facility, zirconium-capped LaBr\(_3\) scintillators, known as their Real Time Neutron Activation Detectors (RT-NADs), register \(^{90}\mathrm{Zr(n,2n)}{}^{89}\mathrm{Zr}\) gamma lines via time resolved spectroscopy to return DT-selective yields within hours rather than days-weeks~\cite{Edwards2016}. More recent, yttrium-capped LaBr\(_3\) detectors (PANDA-FES) use the same system with a different cap exploiting \(^{79}\)Br and \(^{89}\)Y activation lines for portable yield metrology on D--D Z-pinch platforms~\cite{Youmans2024}. While the integration of the activation material with high resolution gamma spectroscopy offers greater selectivity, LaBr\(_3\) crystals and the integrated digitizer bases used are relatively expensive and the electronics can suffer upsets or failures under high D--T fluences, complicating routine operation near powerful sources.

Cherenkov approaches require no scintillating material or wavelength shifter and provide inherent rejection of sub-threshold radiation, simplifying the detection chain relative to scintillator-based activation detectors. The Cherenkov photon yield in fused silica is roughly three orders of magnitude lower than typical scintillator light output per MeV of electron energy. With the additional advantage that Cherenkov photons being prompt, having no decay time or afterglow, allowing high count rate applications. Water-flow \(^{16}\)N monitors count \(\beta\)-induced Cherenkov light for real-time D–T monitoring in steady-state or long-pulse systems~\cite{Verzilov2004}. In clear, non-scintillating glasses, activation products can be read directly via Cherenkov emission, with multi-exponential fits separating contributors by half-life; such glass/Cherenkov methods have even been explored for nuclear-detonation forensics~\cite{Bell2010,Peplow2014}. 

Undoped fused silica (SiO\(_2\)) offers a simpler, inert, non-toxic alternative that integrates activation target and radiator: short-lived \(^{28}\)Al and \(^{16}\)N \(\beta^-\) decays produced by \(^{28}\mathrm{Si(n,p)}{}^{28}\mathrm{Al}\) and \(^{16}\mathrm{O(n,p)}{}^{16}\mathrm{N}\) generate Cherenkov light in the same material. The result is a compact, D–T-selective diagnostic without Be/As hazards or LaBr-based cost/fragility, and with straightforward optical readout.

\section{Detector and Readout}
\subsection{Fused Silica Optics}
The sensor is a high-purity fused-silica (SiO\textsubscript{2}) rod, \SI{6}{in} long by \SI{1}{in} in diameter that has been end-polished and wrapped in eight layers of PTFE reflector. One end of the rod is optically coupled to a Hamamatsu H10580 photomultiplier tube (PMT) assembly that includes a partially active voltage divider base for improved saturation characteristics~\cite{Hamamatsu-Catalog}. The coupling was first made with silicone optical grease and later with EPO-TEK~301 clear epoxy for long-term stability~\cite{Epotek301}. The assembly is made light-tight with a black 3D-printed PLA cover; no scintillator or wavelength shifter is used (Fig.~\ref{fig:detector}).

\subsection{Activation channels}
\begin{figure}[t]
\centering
\includegraphics[width=\columnwidth]{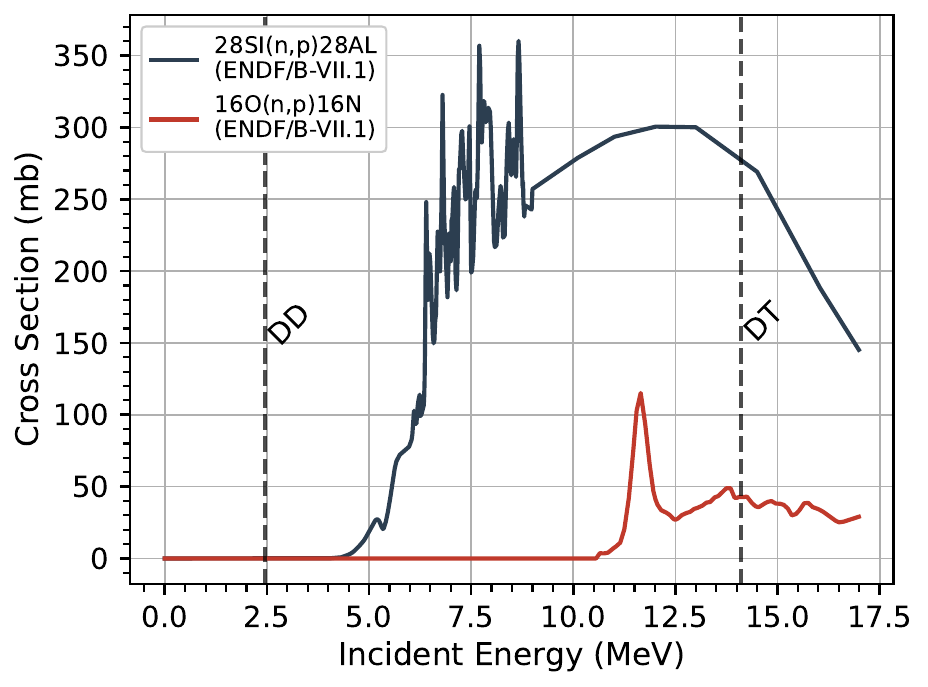}
\caption{ENDF/B-VII.1 evaluated cross sections vs incident neutron energy for \(^{28}\mathrm{Si(n,p)}^{28}\mathrm{Al}\) and \(^{16}\mathrm{O(n,p)}^{16}\mathrm{N}\), plotted using the Curie toolkit~\cite{CurieLib}.}
\label{fig:cs_endf}
\end{figure}

Primary D--T-relevant activation channels in fused silica are summarized in Table~\ref{tab:activation}. Evaluated cross sections and thresholds are taken from ENDF/B-VII.1~\cite{Chadwick2011} (see Fig.~\ref{fig:cs_endf}).

\begin{table*}[t]
\caption{Activation channels in fused silica (Si and O targets). Thresholds are reaction thresholds; endpoint energies are approximate ~\cite{Chadwick2011}.}
\label{tab:activation}
\centering
\begin{tabular}{@{}lllll@{}}
\toprule
Target (abundance) & Reaction & Threshold (MeV) & Product half-life & \(\beta^-\) endpoint (MeV) \\
\midrule
\textsuperscript{28}Si (92.2\%) & \(^{28}\mathrm{Si(n,p)}^{28}\mathrm{Al}\) & \(\sim4.6\) & \SI{134}{s} & \(\sim2.86\) \\
\textsuperscript{16}O (99.8\%)  & \(^{16}\mathrm{O(n,p)}^{16}\mathrm{N}\)  & \(\sim10\) & \SI{7.13}{s} & \(\sim10.4\) \\
\bottomrule
\end{tabular}
\end{table*}

With the refractive index of fused silica being \(n \approx 1.47\) at 400\,nm, the Cherenkov condition is \(\beta_v > 1/n,\) where \(\beta_v = v/c\), \(v\) is the electron velocity, and \(c\) is the speed of light in vacuum. This gives a threshold \(\beta_v \approx 0.685\), corresponding to an electron kinetic energy of about \(186\ \text{keV}\).

Because both the \(^{16}\)N (\(\sim 10.4\) MeV) and \(^{28}\)Al (\(\sim 2.86\) MeV) \(\beta^-\) endpoints far exceed the 186 keV Cherenkov threshold, the vast majority of decay electrons produce Cherenkov light. Although this emission is concentrated in the UV--blue spectrum, higher-energy electrons generate disproportionately more photons per track, resulting in high detection-weighted efficiency.

Above this threshold, the number of Cherenkov photons emitted per unit path length and per unit wavelength interval depends on the electron velocity and the refractive index of the medium. Neglecting constants, the differential yield can be expressed as
\begin{equation}
\frac{\mathrm{d}^2 N}{\mathrm{d}x\,\mathrm{d}\lambda}
\propto \frac{1}{\lambda^{2}}
\left( 1 - \frac{1}{\beta_v^{2} n^{2}(\lambda)} \right),
\qquad \beta_v > \frac{1}{n(\lambda)} \, .
\end{equation}
Here, \(x\) is the track length and \(\lambda\) the photon wavelength. The term \(1/\lambda^2\) favors shorter wavelengths (blue/UV), while the factor \(\bigl(1 - 1/\beta_v^{2} n^{2}\bigr)\) increases as the electron velocity rises above threshold. Thus, higher-energy electrons generate more Cherenkov photons over the detectable band, with the total yield scaling approximately with both track length and electron energy. The high energy electrons from the \(\beta^-\) decays of \(^{16}\mathrm{N}\) and \(^{28}\mathrm{Al}\) produce detectable light along multi-centimeter tracks in the fused-silica rod~\cite{PDG2012}.

\subsection{Electronics}
The detector was fielded over \SI{82}{ft} of RG223 \SI{50}{\ohm} coaxial cable to emulate the Polaris installation. The PMT was biased to \(-\SI{1300}{V}\) and the anode signal digitized by a CAEN DT5730 digitizer (14-bit, \SI{500}{MS/s}) running DPP-PSD firmware in list-mode~\cite{CAEN-DT5730}. A leading-edge threshold above the PMT noise floor registers an event; after which the firmware timestamps and integrates each pulse over short and long windows. Unless otherwise stated the threshold was set to \SI{22.9}{mV}; thresholds of 61, 122, and \SI{244}{mV} were also used to reduce prompt decay count rates and therefore increase the measurable dynamic range for a given maximum throughput of the digitizer.

\begin{figure*}[t]
\centering
\begin{subfigure}[t]{0.5\textwidth}
  \centering
  \includegraphics[width=\linewidth,height=0.28\textheight,keepaspectratio]{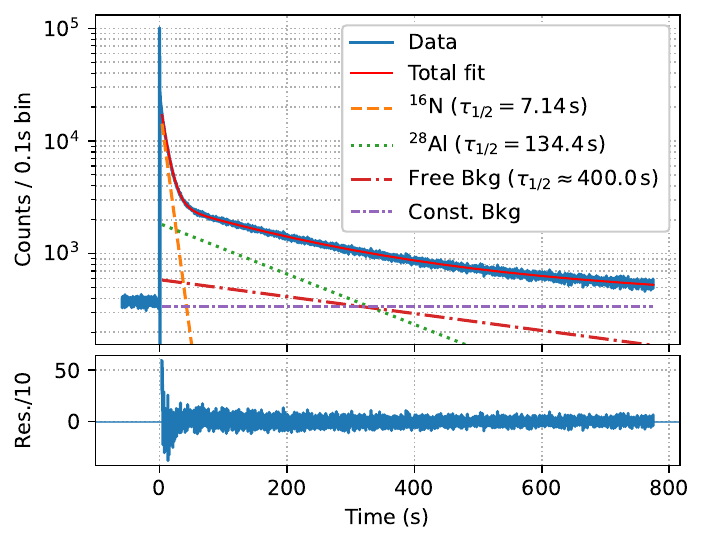}
  \caption{}
\end{subfigure}\hfill
\begin{subfigure}[t]{0.5\textwidth}
  \centering
  \includegraphics[width=\linewidth,height=0.28\textheight,keepaspectratio]{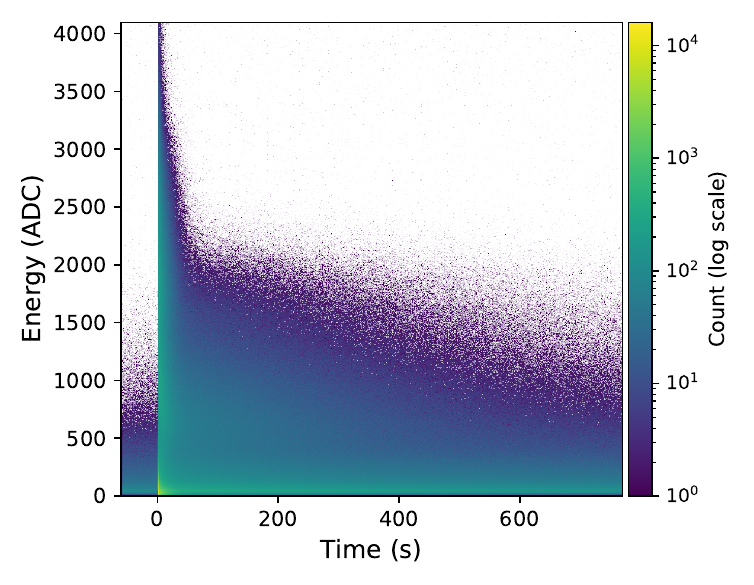}
  \caption{}
\end{subfigure}
\caption{(a) ZEUS post-shot count rate \(R(t)\); points: \SI{0.1}{s} bins for shot 241022-06; line: best-fit model with individual components (fit starts at \(t=\SI{1}{s}\) to exclude PMT/digitizer saturation). (b) 2D histogram of per-event integral (energy proxy) vs time.}
\label{fig:zeus_panel}
\end{figure*}

\section{Acquisition and Analysis}
\subsection{Binning and model}
Events are accumulated for \(>\SI{600}{s}\) after each shot and binned in \SI{0.1}{s} intervals to form a rate history \(R(t)\). For data \SI{1}{s} past the trigger, we fit
\begin{equation}
R(t) = N_0^{(^{16}\mathrm{N})} e^{-t/\tau_1}
     + N_0^{(^{28}\mathrm{Al})} e^{-t/\tau_2}
     + N_0^{(\mathrm{bkg})} e^{-t/\tau_3}
     + C ,
\end{equation}
with \(\tau_1=T_{1/2}^{(^{16}\mathrm{N})}/\ln 2=\SI{10.29}{s}\), \(\tau_2=T_{1/2}^{(^{28}\mathrm{Al})}/\ln 2=\SI{193.4}{s}\), \(\tau_3\) free (bounded \(\geq\SI{400}{s}\)), and \(C\) constant, similar to that used for other glass-Cherenkov activation analyses~\cite{Bell2010,Peplow2014}. An example of this bin and fit can be found in Figure \ref{fig:zeus_panel}.
The free exponential term (bounded \(\tau_3 \ge \SI{400}{s}\)) is included to capture longer-lived contributions such as activation of trace impurities in the fused silica, activation of surrounding structural materials (PTFE reflector, PLA housing), or residual PMT phosphorescence. The constant \(C\) captures the PMT dark count rate and slow background radiation.
For a typical ZEUS D--T shot, the \(^{16}\)N count rate at \(t=\SI{1}{s}\) exceeds the fitted background \((C\) plus the slow exponential\()\) by roughly two orders of magnitude, and the \(^{28}\)Al component remains above background for approximately \SI{400}{s}, providing signal-to-background ratios well suited to the fit.

\subsection{D--D and D--T Testing}
At the Z-pinch Experimental Underground System (ZEUS) Dense Plasma Focus (DPF) located at the Nevada National Security Site~\cite{NNSS2025}, the SiO\textsubscript{2} detector was fielded directly above the pinch at a radial distance of \(r=\SI{21.6}{cm}\). The detector was calibrated against the facility’s silver activation detector, which is located in an EMI cabinet at a fixed position to the side of the pinch and is regularly calibrated against a praseodymium activation puck. The Praseodymium puck is exposed to the fluence from a DPF shot and is then counted on a NaI coincidence counter calibrated against a NIST traceable \(^{22}\mathrm{Na}\) source. For each shot an independent yield \(Y_{\mathrm{cal}}\) and the detector standoff give the fluence \(\Phi\), setting \(N_0 = K\,\Phi\). Where \(K\) is the calibration constant defined for \(^{16}\mathrm{N}\) and \(^{28}\mathrm{Al}\) independently. Multiple shots established linearity.

To verify insensitivity to D--D neutrons, the detector was also tested at the MegaJOuLe Neutron Imaging Radiography (MJOLNIR) D--D DPF at the Lawrence Livermore National Laboratory under the same DAQ settings ~\cite{Schmidt2021}.

\subsection{Uncertainties}
The fused-silica detector is operated as a precision relative-yield monitor. Accordingly, the intrinsic uncertainty is dominated by counting statistics and the multi-exponential fit, while the absolute-scale uncertainty is calibration-specific and enters as an overall normalization.

The multi-exponential fit (Eq.~2) is performed via nonlinear least-squares minimization on Poisson-weighted residuals. With \(\tau_1\) and \(\tau_2\) fixed, the free parameters are \(N_0(^{16}\mathrm{N})\), \(N_0(^{28}\mathrm{Al})\), \(N_0(\mathrm{bkg})\), \(\tau_3\), and \(C\). The statistical uncertainty for each component is extracted from the fit covariance matrix. Because the half-lives of \(^{16}\mathrm{N}\) and \(^{28}\mathrm{Al}\) differ by a factor of \(\sim 19\), temporal separation is excellent, resulting in only modest correlation between the primary amplitudes (typically \(|\rho|<0.3\), where \(\rho\) is the correlation coefficient from the fit covariance, in our datasets). Time-bin quantization introduces a bounded start-time error that is absorbed primarily by the fitted amplitudes (sub-\SI{1}{\percent} effect for \(^{16}\mathrm{N}/^{28}\mathrm{Al}\) at \SI{0.1}{s} bins). Geometric standoff uncertainty enters as a separate scale factor on the inferred fluence.

For absolute yield reporting, the detector relies on cross-calibration constants (\(K_{\mathrm{Al}}\), \(K_{\mathrm{N}}\)) derived from the ZEUS facility reference or Polaris neutron activation foil measurements. This introduces a systematic uncertainty inherited from the reference diagnostic (e.g., praseodymium calibration, reference geometry, and plasma anisotropy). Because a full systematic error budget for the ZEUS facility yield was not available, the calibration constants are scaled to the nominal reported yields; thus the uncertainties reported here reflect only the intrinsic statistical and fitting precision of the SiO\textsubscript{2} detector, while the absolute systematic uncertainty acts as a global scaling factor on the final fluence.

\section{Results}
\subsection{ZEUS D--T validation and calibration}
\begin{figure}[b]
\centering
\includegraphics[width=\columnwidth]{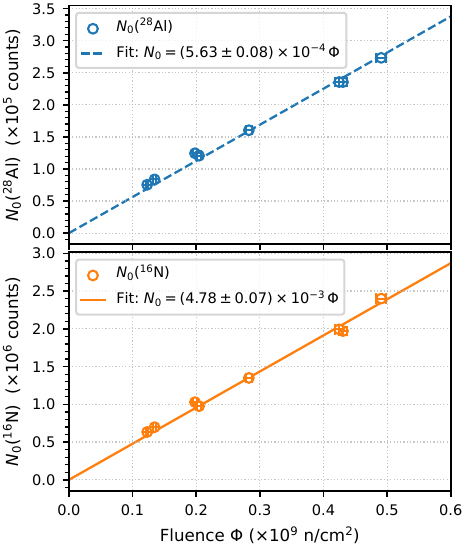}
\caption{Linearity of fitted activation components vs ZEUS fluence. Calibrated \(K_{\mathrm{Si}\to\mathrm{Al}}\) and \(K_{\mathrm{O}\to\mathrm{N}}\) against an independent ZEUS yield-to-fluence reference.}
\label{fig:linearity}
\end{figure}

Detector operated at the radial distance, negative high voltage, and digitizer thresholds as specified above. Data was collected for approximately \(\SI{60}{s}\) prior to the shot and \(\SI{700}{s}\) following the shot. Figure~\ref{fig:zeus_panel}a shows the binned rates \(R(t)\) with \(\SI{7.14}{s}\), \(\SI{134.4}{s}\),  and \(\geq\SI{400}{s}\) components fitted. A set of reference shots of different yields establish linearity between the Silver detector derived facility fluences and both components of the SiO\textsubscript{2} activation detector. The fit-derived \(N_0^{(^{16}\mathrm{N})}\) and \(N_0^{(^{28}\mathrm{Al})}\) were plotted and fitted calibration constants of $K_{\mathrm{Al}} = (5.63 \pm 0.08)\times 10^{-4} \, \text{counts}\cdot\text{cm}^2/\text{n}$ and $K_{\mathrm{N}} = (4.78 \pm 0.07)\times 10^{-3} \, \text{counts}\cdot\text{cm}^2/\text{n}$ (Figure \ref{fig:linearity}). The precise temporal decomposition into \(\SI{7.13}{s}\) and \(\SI{134}{s}\) components, matching known nuclear data, confirms the signal arises from the claimed activation channels and rules out alternative explanations for the linear trend, such as PMT afterpulsing or slow phosphorescence. In Figure \ref{fig:linearity}, the error bars are derived from the uncertainty of the fitting process.

\subsection{MJOLNIR D--D insensitivity}
\begin{figure}[t]
\centering
\includegraphics[width=\columnwidth]{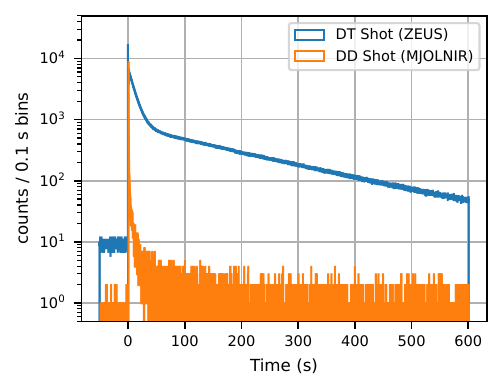}
\caption{Comparison of ZEUS (D--T) and MJOLNIR (D--D) shots at comparable fluence. Post-shot \(R(t)\) shows minimal signal for D--D neutrons.}
\label{fig:mjolnir_rt}
\end{figure}

The detector operated near the MJOLNIR D--D DPF under the same DAQ settings as specified above. By placing the detector closer to the pinch, it was exposed to a nominal D--D fluence comparable to the lower-end ZEUS D--T shots. No count rate increase was observed besides the prompt detection of the shot (Fig.~\ref{fig:mjolnir_rt}). This null result is attributed to neutron energy rather than fluence: the ENDF/B-VII.1 cross sections for \(^{28}\mathrm{Si(n,p)}\) and \(^{16}\mathrm{O(n,p)}\) are effectively zero at \SI{2.45}{MeV} (Fig.~\ref{fig:cs_endf}), and applying the ZEUS-derived calibration constants shows that \SI{14.1}{MeV} neutrons at this nominal fluence would produce a detectable signal.
Dense plasma focus devices operating with deuterium produce secondary 14.1~MeV neutrons via burn-up of approximately 1~MeV tritons from the D(d,p)T reaction branch. Triton burn-up is a standard feature of deuterium fusion plasmas and produces a secondary neutron population whose magnitude depends on triton confinement and plasma density \cite{Huba2016,Stacey2010,Nishitani2021}. In magnetically confined plasmas, $Y_{DT}/Y_{DD}$ can reach $10^{-3}$--$10^{-2}$. In DPF devices, however, the plasma is short-lived, spatially small, and weakly confining for 1~MeV tritons, so escape losses are expected to strongly suppress burn-up. The resulting secondary 14.1~MeV yield should therefore be well below $10^{-3}$ of the primary D--D yield and does not impose a practical lower bound on the diagnostic sensitivity considered here.

\section{Discussion}
\subsection{Operating envelope}
\begin{figure}[t]
\centering
\includegraphics[width=\columnwidth]{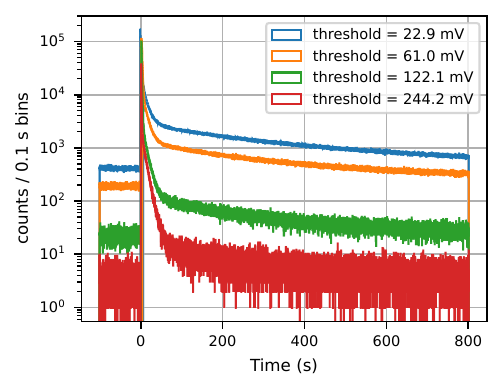}
\caption{Effect of digitizer threshold on \(R(t)\). Increasing the threshold reduces sensitivity to \(^{28}\mathrm{Al}\) decays and backgrounds, making the \(^{16}\mathrm{N}\) component more prominent.}
\label{fig:thresholds}
\end{figure}

\begin{figure}[t]
\centering
\includegraphics[width=\columnwidth]{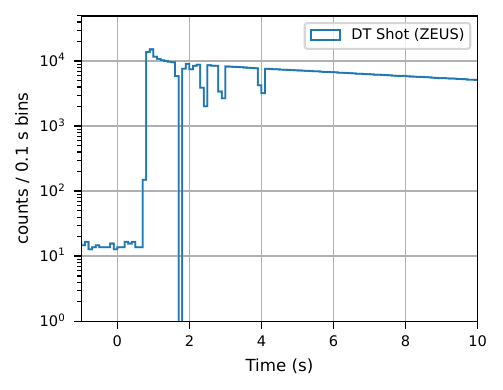}
\caption{Early-time region showing list-mode throughput limitations following a D--T shot.}
\label{fig:saturation}
\end{figure}

The detector provides rapid automated yield readout with minimal operator burden. The practical lower yield is set by counting statistics required for the fit, where fits to the \(^{16}\mathrm{N}\) decay have been observed to converge for fluences of \(\sim10^{2}\) n/cm\(^2\). At high yields, early-time pileup and list-mode throughput limits can cause digitizer dropouts (Fig.~\ref{fig:saturation}); mitigation includes increased standoff, a smaller SiO\textsubscript{2} radiator, higher thresholds, or higher-throughput firmware. From Figure \ref{fig:saturation}, the digitizer max rate is observed to be \(\sim 10^4\) counts/s, as observed by the dropouts in the digitizers sustained output. Raising the digitizer threshold as observed in Figure \ref{fig:thresholds} lowers observed count rate.

\subsection{Selectivity and background}
The activation thresholds (\(\sim\SI{4.6}{MeV}\) for \(^{28}\)Si and \(\sim\SI{10}{MeV}\) for \(^{16}\)O) render the detector insensitive to D--D fusion neutrons (\SI{2.45}{MeV}) and most room-return or moderated neutrons. In a D--T fusion environment, the dominant above-threshold source is the \SI{14.1}{MeV} fusion neutron; other potential contributors (e.g., breakup or scattered neutrons above threshold) would require separate accounting if present at significant levels. Internal \(\gamma\) rays from the same activation chains contribute a smaller Cherenkov component via Compton electrons with the same half-life signatures.
Because the \(^{28}\mathrm{Si(n,p)}\) and \(^{16}\mathrm{O(n,p)}\) reactions have different energy thresholds (\(\sim\)\SI{4.6}{MeV} and \(\sim\)\SI{10}{MeV}, respectively), the ratio of the fitted \(^{28}\)Al and \(^{16}\)N amplitudes is sensitive to the neutron energy spectrum. A pure \SI{14.1}{MeV} source produces a fixed ratio set by the respective cross sections and detection efficiencies;  a down scattered-neutron component in the \SIrange{4.6}{10}{MeV} range would preferentially boost the \(^{28}\)Al component. Monitoring this ratio can provide spectroscopic information about neutron energy distributions.

\subsection{Comparison to alternatives}
Classical activation foils (In, Cu, Zr, etc.) achieve more robust measurements but require extraction and extended measurements on dedicated setups; the fused silica activation detector is remotely read out and intrinsically D--T selective. Compared to NIF’s RT-NADs (LaBr\textsubscript{3})~\cite{Edwards2016}, the fused silica approach is lower cost and easily deployable. Relative to prompt scintillation detectors, the activation–Cherenkov method integrates fluence into a decay signal that is read out seconds to minutes after the pulse, inherently avoiding the prompt EMI transient associated with pulsed fusion generators (the fit window begins at \(t=\SI{1}{s}\); see Section 4.1). Water-based \(^{16}\)N Cherenkov monitors are ideal for steady/long pulses~\cite{Verzilov2004}; a solid fused-silica detector is optimized for discrete pulses and requires no fluid handling. Our undoped fused-silica approach leverages multi-exponential Cherenkov counting demonstrated in glass systems~\cite{Bell2010,Peplow2014} while avoiding toxic or scarce activation materials.

\subsection{Radiation effects and reliability}
High radiation doses can induce color centers in silica, increasing optical attenuation~\cite{Girard2013}. We plan to track the fused silica detector's optical transmission, PMT gain, and dark rate under higher-yield operation. Periodic Zr-foil activations will continue to anchor absolute calibration. If needed, the diagnostic can easily be replaced given the low cost. The H10580 PMT's susceptibility to radiation will also be monitored where the borosilicate optical window can also darken and the diodes used in the partially active voltage divider could degrade in performance. Future iterations of this detector would look at optimizing the volume and geometry of the fused silica-PMT assembly and also make use of radiation tolerant PMTs, replacing the use of the active voltage divider with a gating mechanism and the borosilicate window with a fused silica one.

\section{Conclusions}
A fused-silica activation Cherenkov detector provides a simple, fast, precise, and D--T-selective relative neutron-yield diagnostic. It delivers pulse-to-pulse relative yields within \(\sim\)\SI{3}{min} using list-mode acquisition and multi-exponential fitting. Measured relative yields agree with a calibrated silver activation detector, and the system is insensitive to D--D neutrons. Its compactness, simplicity, and low cost make it suitable for scaling to multiple channels or deployment at smaller facilities. Ongoing Polaris operation will expose the detector to higher fluences and allow evaluation of durability and calibration stability against standard Zr foils.

\section*{Acknowledgments}
Part of this work was performed at the Z-pinch Experimental Underground System (ZEUS). We appreciate the assistance of Daniel Lowe, Adam Ikehara, Megan Lawrence, Rafael Castro and Giovani Acevedo for facility access and D--T fluence references ~\cite{NNSS2025}.

Part of this work was performed at the MegaJOuLe Neutron Imaging Radiography (MJOLNIR) Dense Plasma Focus. We appreciate the assistance of Amanda Youmans and Drew Higginson for facilitating access and D--D reference fluences ~\cite{Schmidt2021}.


\end{document}